\documentclass[aoas,preprint]{imsart}

\def\bSig\mathbf{\Sigma}

\usepackage[figuresright]{rotating}
\usepackage{graphicx}
\usepackage{subfigure}
\usepackage{amsmath}
\usepackage{hyperref}
\usepackage{multirow}
\usepackage{algorithm}
\usepackage{algorithmic}
\usepackage{enumerate}
\usepackage{csvsimple}
\usepackage{datatool}
\usepackage{booktabs}
\usepackage{color}
\usepackage[anythingbreaks]{breakurl}
\usepackage{amsfonts}
\usepackage{thmtools}
\usepackage{amssymb,bbm,amsthm,amsmath}
\usepackage[normalem]{ulem}

\makeatletter

\newcommand{\Rmnum}[1]{\expandafter\@slowromancap\romannumeral #1@}
\makeatother
\RequirePackage[OT1]{fontenc}
\RequirePackage{amsthm,amsmath}
\RequirePackage[authoryear]{natbib}
\RequirePackage[colorlinks,citecolor=blue,urlcolor=blue]{hyperref}

\arxiv{arXiv:0000.0000}

\startlocaldefs
\numberwithin{equation}{section}
\theoremstyle{plain}

\newtheorem*{remark}{Remark}
\endlocaldefs

\newcommand{\HS}{\mathrm{HS}}

\newcommand{\sgn}{\text{sign}}

\begin{document}

\begin{frontmatter}
\title{
P-value evaluation, variability index and biomarker categorization for adaptively weighted Fisher's meta-analysis method in omics applications\thanksref{T1}
}
\runtitle{Issues in AW-Fisher}

\begin{aug}
\author[1]{\fnms{Zhiguang} \snm{Huo}\ead[label=e1]{zhuo@ufl.edu}},
\author[3]{\fnms{Shaowu} \snm{Tang}\ead[label=e3]{shaowu.tang@roche.com}},
\author[2]{\fnms{Yongseok} \snm{Park}\ead[label=e2]{yongpark@pitt.edu}},\thanksref{co}
\and
\author[2]{\fnms{George} \snm{Tseng}\ead[label=e4]{ctseng@pitt.edu}}$^,$\thanksref{co}

\thankstext{co}{To whom correspondence should be addressed.}
\thankstext{T1}{Supported by NIH R01CA190766 for Z.H. and G.T.}

\runauthor{Z. Huo et al.}

\affiliation[1]{University of Florida} 
\affiliation[2]{University of Pittsburgh} \and 
\affiliation[3]{Roche Molecular Systems, Inc}

\address{Zhiguang Huo\\
Department of Biostatistics\\ 
University of Florida \\
Gainesville, FL 32611\\
\printead{e1}\\
\phantom{E-mail:\ }}

\address{Yongseok Park\\
Department of Biostatistics
University of Pittsburgh \\
Pittsburgh, PA 15261\\
\printead{e2}\\
}

\address{Shaowu Tang\\
Roche Molecular Systems, Inc\\
San Francisco, CA  94588\\
\printead{e3}\\
}

\address{George Tseng\\
Department of Biostatistics, Human Genetics \\ 
and Computational Biology\\
University of Pittsburgh \\
Pittsburgh, PA 15261\\
\printead{e4}\\
}
\end{aug}

\begin{abstract}

Meta-analysis methods have been widely used to combine results from multiple clinical or genomic studies to increase statistical power
and ensure robust and accurate conclusion.
Adaptively weighted Fisher's method (AW-Fisher) is an effective approach to combine p-values from $K$ independent studies
and to provide better biological interpretation by characterizing which studies contribute to meta-analysis.
Currently, AW-Fisher suffers from lack of fast, accurate p-value computation and variability estimate of AW weights.
When the number of studies $K$ is large,
the $3^K - 1$ possible differential expression pattern categories can become intractable.
In this paper, we apply an importance sampling technique with spline interpolation to increase accuracy and speed of p-value calculation.
Using resampling techniques, 
we propose a variability index for the AW weight estimator and a co-membership matrix to characterize pattern similarities between genes.
The co-membership matrix is further used to categorize differentially expressed genes based on their meta-patterns for further biological investigation.
The superior performance of the proposed methods is shown in simulations.
These methods are also applied to two real applications 
to demonstrate intriguing biological findings. 
\end{abstract}

\begin{keyword}
\kwd{adaptive weights}
\kwd{Fisher's method}
\kwd{meta-analysis}
\kwd{differential expression analysis}
\end{keyword}

\end{frontmatter}


\section{Introduction}
\label{sec:intro}

High-throughput biological experiments play a key role in deciphering biological mechanisms behind complex diseases.
Advanced experimental techniques allow us to obtain high-resolution genomic information with affordable price.
Over the years large amount of omics data are accumulated in public databases and depositories:
The Cancer Genome Atlas (TCGA) \url{http://cancergenome.nih.gov}, 
Gene Expression Omnibus (GEO) \url{http://www.ncbi.nlm.nih.gov/geo/} and
Sequence Read Archive (SRA) \url{http://www.ncbi.nlm.nih.gov/sra/}, just to name a few.
For a given transcriptomic study from microarray or RNA-seq,
many statistical methods have been developed for detecting differentially expressed (DE) genes as candidate biomarkers \citep{pan2002comparative, soneson2013comparison}.
The analysis of single study, however, contains small to moderate sample size (usually  $N = 20 \sim$ 50), 
producing unstable and inaccurate results \citep{simon2003pitfalls,simon2005development,domany2014using}.
Meta-analysis to combine multiple transcriptomic studies has become a common practice to improve statistical power and reproducibility.
Interested readers may refer to \cite{ramasamy2008key} for a practical guideline of microarray  meta-analysis, 
and \cite{tseng2012comprehensive, begum2012comprehensive} for comprehensive reviews of microarray and genome-wide association study (GWAS) meta-analysis.

Among the numerous meta-analysis methods proposed in the literature,
combining p-values from multiple studies is a simple and flexible solution to combine studies of different experimental design 
and avoid complexity from batch effect (e.g. different studies utilize different platforms or experimental protocols).
Multiple hypothesis settings  have been considered to address different biological questions.
According to  \cite{song2014hypothesis} (see also \cite{birnbaum1954combining, li2011adaptively}), 
three major hypothesis settings have been considered in the literature: 
$\HS_A$ 
targets on detecting biomarkers that are differentially expressed in all cohorts ($H_0: \vec{\boldsymbol{\theta}}\in \bigcap \{ \theta_k=0\}$
vs $H_A: \vec{\boldsymbol{\theta}}\in \bigcap \{ \theta_k \ne 0\}$,
where $\theta_k$ is the effect size of study $k$, 
$1\le k \le K$);
$\HS_B$ 
targets on biomarkers differentially expressed in one or more studies
($H_0: \vec{\boldsymbol{\theta}}\in \bigcap \{ \theta_k=0 \}$ vs 
$H_A: \vec{\boldsymbol{\theta}}\in \bigcup \{ \theta_k \ne 0 \}$);
$\HS_r$ 
targets on biomarkers differentially expressed in at least $r$ studies
($H_0: \vec{\boldsymbol{\theta}}\in \bigcap \{ \theta_k=0 \}$ vs 
$H_A: \sum \mathbb{I} { \{ \theta_k \ne 0 \} } \ge r$,
where $\mathbb{I} { \{ \cdot \} } $ is an indicator function taking value one if the statement is true and zero otherwise 
and $r$ is usually pre-specified with $K/2 \le r \le K$).
Biologically $\HS_A$ is preferred when the purpose is to find concordant genes across all studies. 
$\HS_r$ can be considered as a robust form of $\HS_A$ to seek for concordant genes in majority of studies.
On the other hand,
$\HS_B$ is considered when heterogeneity is expected and we are interested in biomarkers statistically significant in at least one study.

In the literature, $\HS_B$ is a union-intersection test (UIT, \cite{roy1953heuristic}) 
and is also called a conjunction or intersection hypothesis \citep{benjamini2008screening}.
Many statistical tests have been developed for this hypothesis setting, including Fisher's method 
\citep{fisher1934statistical}, Stouffer's \citep{stouffer1949american} method,
minimum p-value method \citep{tippett1931methods} and many others.
Fisher's method defines the test statistic by sum of log-transformed p-values: 
$T^{F} = -2\sum_{k=1}^K \log p_k$, where $p_k$ is the p-value from the $k^{th}$ study;
Stouffer's method uses $T^{S} = - \frac{1}{\sqrt{K}} \sum_{k=1}^K \Phi^{-1}(p_k)$ 
where $\Phi^{-1}(\cdot)$ is the inverse CDF of standard normal distribution.
A larger Fisher (or Stouffer)  score indicates stronger differential expression evidence.
Under the null assumption and assuming independence across studies,
the null distribution of Fisher's statistics follows $\chi^2_{2K}$ and Stouffer's follows $N(0,1)$.
Although Fisher's method has many theoretical advantages 
(e.g.  asymptotic Bahadur optimality under certain restricted Gaussian assumptions; see \cite{littell1971asymptotic}), 
it has a critical pitfall when heterogeneity is expected across studies.
For example, 
suppose $\vec{p}_1 = (0.001,1,1)$ represents p-values of three studies of gene 1 
and $\vec{p}_2 = (0.1,0.1,0.1)$ represents p-values of gene 2. 
Both genes produce the same Fisher's test statistics and meta-analysis p-values
($T^{F} = 13.8$ and $p^{F}=0.032$) but the biological interpretations of the two genes are obviously different.
$\vec{p}_1$ indicates strong statistical significance only in the first study,
while $\vec{p}_2$ shows marginal statistical significance in all three studies.
To characterize study heterogeneity in meta-analysis, \citet{li2011adaptively} proposed an adaptively weighted Fisher's method (AW-Fisher) 
where the Fisher's score is modified as weighted sum and the 0-1 weights can be viewed as latent variable of whether a study contributes DE information to the meta-analysis (details see Section~\ref{sec:awFisher}).
Aside from additional biological interpretation of AW weights, AW-Fisher also enjoys nice theoretical properties. 
It has been shown to be admissible \citep{li2011adaptively} and asymptotic Bahadur optimal under certain Gaussian assumptions \citep{Park2017AWtheory}.
In addition, Fisher's method is more powerful when all studies are significant and 
minimum p-value method is more powerful when only one study has small p-value.
AW-Fisher theoretically takes advantage of both  methods on their favored extreme situations \citep{li2011adaptively}.
\cite{chang2013meta} performed a comprehensive comparative study to evaluate 12 popular microarray meta-analysis methods 
and categorized them into the three complementary hypothesis settings, $\HS_A$, $\HS_B$ and $\HS_r$.
AW-Fisher was the best performer in the $\HS_B$ setting when considering a variety of data and heterogeneity assumptions.

Despite practical and theoretical advantages of AW-Fisher, 
currently there exist three major issues when applying the method. 
Firstly, p-value calculation of AW-Fisher has no simple closed-form solution. 
Permutation analysis is slow and generates low numerical precision of p-values to effectively account for multiple comparisons when thousands of genes are tested simultaneously \citep{sun2010geometric}.  
Secondly, the weight estimate for AW-Fisher is a hard classification (i.e. decision of 0 or 1) and is lack of a variability estimate of the weight. 
Finally, when number of studies $K$ is large, the number of biomarker categories by AW-Fisher weights
 increases exponentially and becomes intractable. 
In this paper, 
we develop methodologies to overcome the three bottlenecks of AW-Fisher. 
In Section~\ref{sec:awFisher}, we introduce AW-Fisher and its existing issues in more detail. 
Section~\ref{sec:fastAW} describes an importance sampling technique with spline interpolation and a linear weight search scheme to overcome computational burden. 
In Section~\ref{sec:variabilityIndex}, we develop a bootstrap scheme to define a variability index of AW-Fisher weight estimate. 
In Section~\ref{sec:clusters}, 
to overcome the exponential growth of number of biomarker categories, we extend the bootstrap scheme to obtain a co-membership matrix to gauge the pattern similarity of resulting biomarkers. 
By applying tight clustering algorithm \citep{tseng2005tight}, tight clusters of biomarkers with different meta-patterns are generated for insightful  biological interpretation and hypothesis generation. 
Section~\ref{sec:real} shows two real applications in mouse metabolism microarray data and HIV transgenic rat RNA-seq data. 
Section~\ref{sec:conclusion} contains final conclusion and discussion.

\section{AW-Fisher and its existing issues}
\label{sec:awFisher}
Below we describe method and rationale for AW-Fisher \citep{li2011adaptively}.
Define $T(\vec{\textbf{P}}; \vec{\textbf{w}} ) = -2 \sum_{k=1}^K w_k \log P_k$,
where $\vec{\textbf{w}} = (w_1, \ldots, w_K) \in {\{ 0,1 \} }^K$ is the AW weight associated with $K$ studies
and $\vec{\textbf{P}} = (P_1, \ldots, P_K) \in {(0,1)}^K$ is the random variable of input p-value vector for $K$ studies.
Under the null distribution and conditional on $\vec{\textbf{w}}$, 
the significance level obtained by $T(\vec{\textbf{P}}; \vec{\textbf{w}} )$ is 
$L(T(\vec{\textbf{P}}; \vec{\textbf{w}} )) = 1 - F_{\chi^2_{d(\vec{\textbf{w}})}}(T(\vec{\textbf{P}}; \vec{\textbf{w}} ))$,
where $d(\vec{\textbf{w}}) = 2\sum_{k=1}^Kw_k$ and
$F_{\chi^2_d}(\cdot)$ is the cumulative distribution function (CDF) of $\chi^2$-distribution  
with degrees of freedom $d$. 
The test statistic of AW-Fisher given p-value vector $\vec{\textbf{P}}$ is defined as 
\begin{equation}
s(\vec{\textbf{P}}) = \min_{\vec {\textbf{w}}} L(T(\vec{\textbf{P}}; \vec{\textbf{w}} ))
\label{eq:AWtestStat}
\end{equation}

\begin{figure}[htbp]
	\centering
		\includegraphics[height=1\columnwidth]{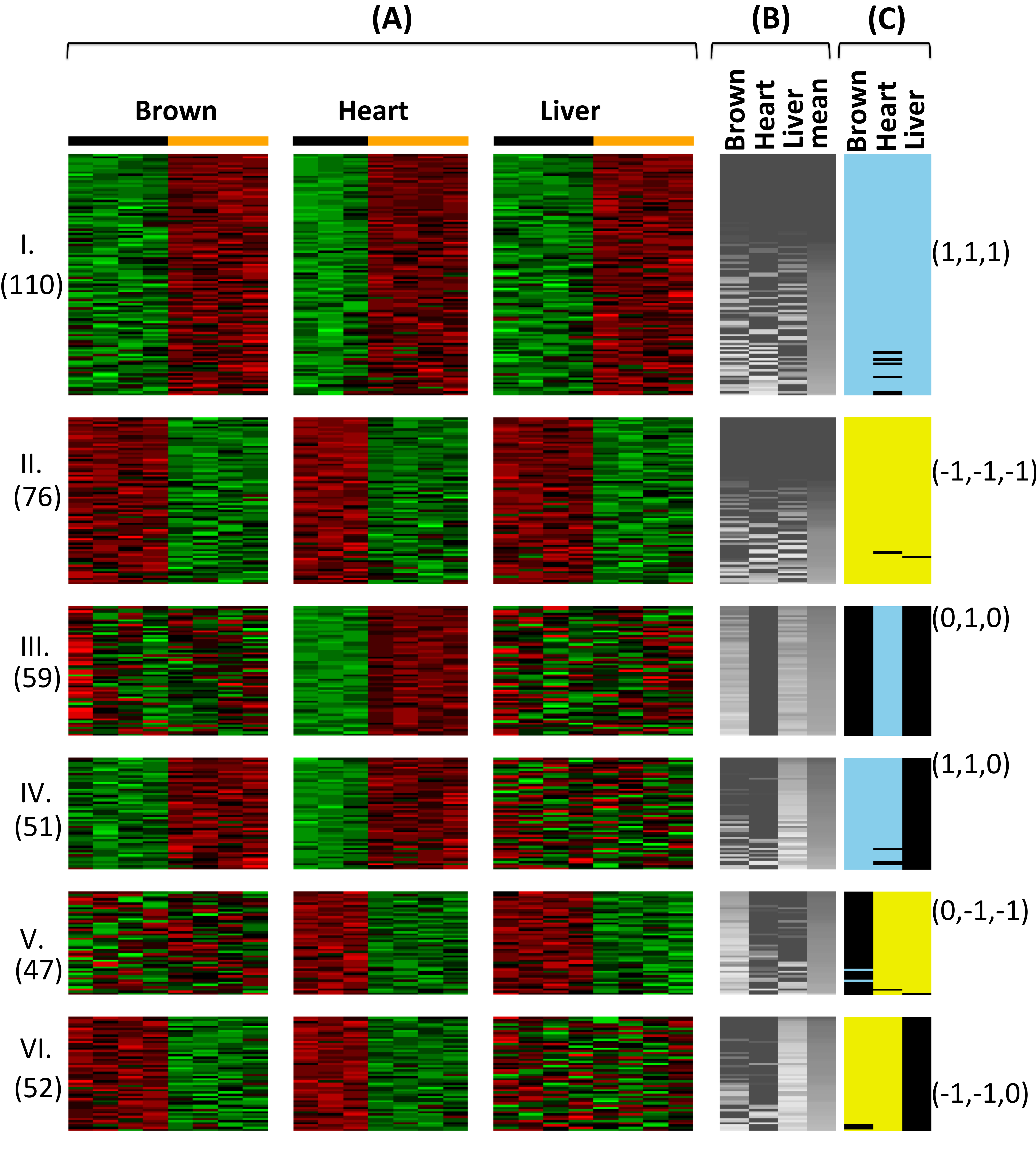}
	\caption{
	Six meta-pattern modules of biomarkers from mouse metabolism example.
	Each gene module (Module \Rmnum{1}, \Rmnum{2}, $\dots$, \Rmnum{6}) shows a set of detected biomarkers with similar meta-pattern of differential signals.
	(A) Heatmaps of detected genes (on the rows) and samples (on the columns) for each tissue (brown fat, heart, liver),
	where each tissue represents a study.
	Black color bar on top represents wild type (control) and orange color bar on top represents VLCAD -/- mice (case).
	Number of genes is shown on the left under each module number. 
	(B) Variability index (genes on the rows and studies on the columns). 
	Variability index is described in Section~\ref{sec:variabilityIndex}. 
	Gray heatmap range from 0 (black) to 1 (white), which is the maximum of the variability index.
	Genes of each module are sorted based on the mean variability index.
	(C) Signed AW-Fisher weights $\hat{v}_{gk}$ for gene $g$ and study $k$.
	Light blue represents $\hat{v}_{gk} = 1$,
	yellow corresponds to $\hat{v}_{gk} = -1$ and black for $\hat{v}_{gk} = 0$.	
	Representative signed AW-Fisher weights for each module are shown on the right.
	Note Brown represents brown fat tissue.
	}
	\label{fig:metabolism_modules}
\end{figure}

The  optimal weight for $\hat{ \textbf{w}}$ is determined by 
$\hat{ \textbf{w}} = w(\vec{\textbf{P}})  = \arg\min_{\vec{\textbf{w}}} L(T(\vec{\textbf{P}}; \vec{\textbf{w}} ))$.
Here we denote by $s$ the mapping from p-value vector to the AW-Fisher test statistic and $S$ is the random variable for AW-Fisher test statistic which can be obtained by $S = s(\vec{\textbf{P}})$.
We further define signed AW-Fisher weights by 
$$\hat{\textbf{v}} = (\hat{v}_1, \ldots, \hat{v}_K) = (\hat{w_1} \cdot \sgn(\hat{\theta}_1), \ldots, \hat{w_K} \cdot \sgn(\hat{\theta}_K)), $$
where $(\hat{\theta}_1, \ldots, \hat{\theta}_K)$ is the estimate of effect size of each study and 
$\sgn (x) = x/|x| $ if $x\ne 0$ and  $\sgn (x) = 0$ otherwise.
Note that $\hat{v}_k$ can be 0, 1 or -1 for $1\le k \le K$.
AW-Fisher is appealing in applications since the AW weight estimate $\hat{\textbf{w}}$ characterizes which study contributes to the meta-analysis result. 
In the previous simple example, 
we have $\hat{\textbf{w}}=(1,0,0)$ for gene 1 and $\hat{\textbf{w}}=(1,1,1)$ for gene 2, which indicates  gene 1 
($\vec{\textbf{P}}=(0.001, 1, 1)$) 
is a first-study-specific biomarker while gene 2 
($\vec{\textbf{P}}=(0.1, 0.1, 0.1)$) 
is an all-study-consistent biomarker. 
Figure~\ref{fig:metabolism_modules}A shows heatmap of candidate biomarkers declared as DE by AW-Fisher's method in a mouse metabolism microarray example  combining three studies (tissues): brown fat, heart, liver (see Section~\ref{sec:metabolism}). 
In each study, VLCAD-/- mutant mice (orange bar on top) were compared to VLCAD+/+ wild-type mice (black bar) and DE analysis was performed using Limma \citep{smyth2005limma}. 
Meta-analysis p-values were calculated for each gene using AW-Fisher method. 
Benjamini-Hochberg's procedure \citep{benjamini1995controlling} was used to account for multiple comparisons and false discovery rate was controlled at 5\% level. 
Among detected biomarkers, some genes are up-regulated DE genes across all tissues (e.g. genes in module \Rmnum{1}, $\hat{v}=(1,1,1)$)); many others are tissue specific 
(e.g. heart-specific biomarkers in module \Rmnum{3}, $\hat{v}=(0,1,0)$).
If applying Fisher's method, these different gene modules will not be distinguished, 
which may hinder biologists for further biological investigation and hypothesis generation. 
Despite the advantages of AW-Fisher in theory and applications,
applying AW-Fisher currently  encounter three major issues outlined below.
    
\begin{enumerate}[I]
\item 
\label{issue:computing}
In the original paper, \citet{li2011adaptively} did not derive a closed-form solution for calculating null distribution of AW statistic. 
Instead,  permutation method (permuting case/control labels in each study independently) was suggested. 
This results in high computing demand, especially high p-value numerical precision is needed to account for multiple comparisons. 
The searching space of all possible weights also becomes high ($2^K-1$) when $K$ goes large. This will limit AW-Fisher in general genomic applications.

\item \label{issue:stable}
The AW weight estimate can generate unexpected discontinuity and is thus not stable.
	For example, the following two genes were taken from the mouse metabolism example in Figure~\ref{fig:metabolism_modules}.
	P-values of the three tissues for probeset $1419484\_a\_at$ were $(0.000391, 0.0962, 0.00211)$,
	and p-values  for probeset $1425567\_a\_at$ were $(0.000356, 0.1026, 0.00206)$.
	Despite their very similar p-value inputs, 
	$1419484\_a\_at$ ended up with AW weight $\hat{\textbf{w}}=(1, 1, 1)$ with p-value $5.64 \times10^{-5}$ using AW-Fisher and
        $1425567\_a\_at$ produced AW weight $\hat{\textbf{w}}=(1, 0, 1)$ with p-value $5.22 \times 10^{-5}$,
        showing unstable weight estimate of the second study.
	In other words, the AW weight estimate is a hard classification with no variability estimate and biomarker categorization is thus unstable.
	
\item 
\label{issue:cluster}
Given $K$ studies, 
the resulting genes could be categorized into $(3^K - 1)$ groups based on their unique AW weight estimate and effect size direction 
(if separating up-regulation and down-regulation into 1 and -1 weight using $\hat{\textbf{v}}$; see Figure~\ref{fig:metabolism_modules}). 
This becomes intractable for further biological investigation when $K$ is large. 
For example, combining $K=5$ studies produces $3^5-1 = 242$ categories of biomarkers.

\end{enumerate}
To solve these issues of AW-Fisher's method, 
we will present methods for fast p-value computing, variability index, biomarker categorization in the following three sections.

\section{Fast computing of AW-Fisher}
\label{sec:fastAW}

In this section we will give solutions to the two computational problems mentioned in Issue~\ref{issue:computing}. 
We propose a fast algorithm of searching the adaptive weights in Section~\ref{sec:linearSearch} 
and an interpolation approach to obtain accurate  p-values in Section~\ref{sec:pValueComputation}.
In Supplementary Section~I, 
we also derive closed-form solution for the cases $K= 2$ to benchmark the performance of the proposed method 
and $K= 3$ for the purpose of demonstrating difficulties of closed-form solution in general $K$.

\subsection{An almost-linear order fast searching algorithm for AW weight $\hat{\textbf{w}}$}
\label{sec:linearSearch}
 
Recall that the searching space $\Omega=\{ \vec{\textbf{w}}: \vec{\textbf{w}} \neq \textbf{0}, \vec{\textbf{w}} = (w_1, \ldots, w_K) \in {\{0,1\}}^K\}$ contains
$2^K-1$ non-zero vectors of weights and searching the whole space $\Omega$
to find 
the AW-Fisher test statistic $s(\vec{\textbf{P}})=\min_{\vec{\textbf{w}} \in\Omega} L(T(\vec{\textbf{P}}; \vec{\textbf{w}} ))$ and the adaptive weights 
$ w(\vec{\textbf{P}}) = \arg\min_{\vec{\textbf{w}} \in\Omega} L(T(\vec{\textbf{P}}; \vec{\textbf{w}} ))$
 becomes computationally expensive when $K$ is large. 
The amount of computation is even more challenging  when the AW-Fisher's method is applied to genomic data, 
where the same procedure is repeated for thousands of genes or even millions of SNPs.  
To overcome this difficulty,
we propose a fast algorithm
to find $\hat{\textbf{w}}$  based on the ordered p-values $\{P_{(i)}\}_{i=1}^{K}$ with $P_{(1)} \le \ldots \le P_{(K)}$.
Specifically,
by decomposing $\Omega$ into  $\Omega=\bigcup_{k=1}^K\Omega_k$ with $\Omega_k=\{ \vec{\textbf{w}}:
\sum_{j=1}^Kw_j=k\}$, 
it can be seen that
$s(\vec{\textbf{P}})
=\min_{\vec{\textbf{w}}\in\Omega}\{ L(T(\vec{\textbf{w}}; \vec{\textbf{P}})) \}
= \min_{ 1\leq k\leq K} \min_{ \vec{\textbf{w}} \in\Omega_k }\{ L(T(\vec{\textbf{w}}; \vec{\textbf{P}})) \}$.
Given $1\leq k_0\leq K$, denote 
by $\vec{\textbf{w}}^{k_0}=(w_1^{k_0},\cdots,w^{k_0}_K)$ the vector of weights such that $-2\sum_{j=1}^Kw_j^{k_0}\log(P_j)=-2\sum_{j=1}^{k_0}\log(P_{(j)})$ 
(i.e. the Fisher's statistics using the first $k_0$ smallest p-values).
Then it is straightforward to see that the test statistic involving the first $k_0$ ordered p-values will generate the most significant $L(T(\vec{\textbf{P}}; \vec{\textbf{w}} ))$ in $\Omega_{k_0}$.
This implies in $\Omega_{k_0}$, only $\vec{\textbf{w}}^{k_0}$ has to be considered for further comparison. 
Therefore, instead of searching the whole space $\Omega$, it is enough to search only $K$ vectors of weights $\{\vec{\textbf{w}}^1,\ldots,\vec{\textbf{w}}^K\}$ to find the adaptive weights $\hat{\textbf{w}}$.
The proposed fast algorithm contains two steps: firstly sorting $K$ p-values (usually with complexity of $\mathcal{O}(K) \log (K)$)
and then searching $K$ vectors of weights (with complexity of $\mathcal{O}(K)$). 
Therefore, the fast searching algorithm proposed in this section reduces the computational complexity from
$\mathcal{O}(2^K)$ to $\mathcal{O}(K \log (K))$, which can significantly reduce computing time when $K$ is large.

\subsection{Importance sampling and interpolation by spline smoothing for fast p-value calculation}
\label{sec:pValueComputation} 

Denote by $\vec{\textbf{p}}_{obs}$ the observed p-values from individual studies and 
$s_{obs} = s(\vec{\textbf{p}}_{obs})$ the observed AW-Fisher statistics.
Theoretically, the p-value of AW-Fisher's method $\mathbbm{P}_{H_0}(S \le s_{obs} )$ can be calculated analytically for any $K\geq 2$.
However, the formulae involves the evaluation of a $K$-fold integral and the integration domain becomes very complicated for
$K\geq 3$, 
which makes the derivation of the closed-form solution tedious and fallible.
For illustration, closed-form derivation of $K=2$ and $K=3$ are shown in Supplementary materials.
In  \cite{li2011adaptively},
a permutation test by randomly permuting class labels in each study was proposed.
Although this non-parametric approach has its merit of maintaining gene dependency structure,
it is computationally demanding and difficult for generating precise small p-value, such as when p-value $< 10^{-4}$,
which is a critical requirement for multiple testing correction on thousands of genes.
In this paper,  
we propose to use importance sampling to obtain an accurate numerical approximation of $\mathbbm{P}_{H_0}(S \le s_{obs} )$.
Importance sampling is a method to accurately estimate expectation of a function with very small value using Monte Carlo sampling method. 
The idea behind importance sampling is to draw samples from a suitable new distribution function
rather than the original one of interest and assign a weight to each sample based on the ratio of two density functions.

To evaluate AW-Fisher p-value $\mathbbm{P}_{H_0}(S \le s_{obs} )$ using importance sampling, 
we propose a beta-distribution density function $f^*(\cdot)$
to draw $\vec{\textbf{P}}$ instead of natural uniform distribution $f(\cdot)$
 so that we can ``over-sample'' those small
p-values that result in a large $S$.
It holds that
\begin{align}
\mathbbm{P}_{H_0}(S \le s_{obs} ) &=\mathbbm{E}_{H_0}[\mathbb{I} \{ S \le s_{obs} \} ]  \label{eq:impoortantSampling}\\
           &= \int \mathbb{I}\{ S \le s_{obs} \} f(\vec{\textbf{P}}) d\vec{\textbf{P}} \nonumber \\
           &= \int \mathbb{I}\{ S \le s_{obs} \} \frac{f(\vec{\textbf{P}})}{f^*(\vec{\textbf{P}})}f^*(\vec{\textbf{P}})d\vec{\textbf{P}} \nonumber \\
           &= \mathbbm{E}^*[\mathbb{I}\{ S \le s_{obs} \} \times W(\vec{\textbf{P}})], \nonumber 
\end{align}
where $f(\cdot)$ is the density of $\vec{\textbf{P}}$ under the null and 
$f^*(\cdot)$ is the proposed density function of $\vec{\textbf{P}}$ for importance sampling.
 Importance sampling weight $W(\cdot) = f(\cdot) / f^*(\cdot)$,
 $\mathbbm{E}(\cdot)$ and $\mathbbm{E}^*(\cdot)$ are 
 the expectation with respect to $f(\cdot)$ and $f^*(\cdot)$ respectively. 
 Therefore, we can obtain expectation from the original measure using a more efficient new one by applying weights for different samples in Monte-Carlo method.
 Under the null hypothesis and independence assumption between different studies, 
 $P_k \sim \mbox{UNIF}(0,1)$ for all $1\le k \le K$, so the joint
distribution of $f(\vec{\textbf{P}}) = 1$.
If we instead use $\text{Beta}(\eta,1)$ distribution as the proposed distribution of each study for importance sampling,
then $f^*(\vec{\textbf{P}})=\eta^K(\prod_{k=1}^K P_k)^{\eta-1}$. 
To implement importance sampling,
suppose we simulate $\vec{p}_i = (p_{i1}, \ldots, p_{iK})$,
where $p_{ik} \overset{i.i.d.}{\sim} \text{Beta}(\eta, 1)$ for $1 \le i \le n$ and $1 \le k \le K$.
Denote by $s_i = s(\vec{p}_i)$. 
From Equation~\ref{eq:impoortantSampling},
we calculate  estimate of $\mathbbm{P}_{H_0}(S \le s_{obs} )$ by
\begin{equation}
\label{eq:impoortantSampling2}
\hat{\mathbbm{P}}_{H_0}(S \le s_{obs}; \eta,  \vec{p}_1, \ldots, \vec{p}_n)
= \frac{1}{n}\cdot\sum_{i=1}^n 
\bigg( 
\mathbb{I}\{s_i \le s_{obs}\} \cdot \frac{1}{\eta^K (\prod_{k=1}^K p_{ik})^{\eta - 1}}
\bigg)
\end{equation}
Our p-value evaluation procedure has the following steps:
\begin{enumerate}
\item Specify targeted $K = 2,3,\ldots, 100$ and targeted AW-Fisher p-values as 
$\{ c_t, t = 1,2,\ldots, 198 \} = \{1,0.99,0.98,0.97,\ldots, 0.03, 0.02, 0.01, 10^{-3}$,
$10^{-4}, 10^{-5}, \ldots, 10^{-100}\}$.

\item 
\label{step:eta}
(Identify suitable $\eta$ for given $c_t$ and $K$) 
Note that different $\eta$ can provide better importance sampling for different range of targeted $c_t$ given $K$.
To identify an appropriate  $\eta$ given $c_t$ and $K$,
we simulate $\vec{q}_i = (q_{i1}, \ldots, q_{iK})$,
where $1 \le i \le 1000$ and $q_{ik} \overset{i.i.d.}{\sim} \text{Unif}(0,1)$.
Denote by $\vec{q}_i^\eta = (q_{i1}^\eta, \ldots, q_{iK}^\eta)$ with element-wise power to  $\eta$
and $r_i^{(\eta)} = s(\vec{q}_i^\eta)$.
Define $r_0 = \text{median}_{1 \le i \le 1000}(r_i^{(\eta)})$.
Note that since $q_{ik} \overset{i.i.d.}{\sim} \text{Unif}(0,1)$,
$q_{ik}^\eta \sim \text{Beta}(\eta,1)$.
From Equation~\ref{eq:impoortantSampling2}, 
we have
\begin{align*} 
\phi(\eta) &=  \hat{\mathbbm{P}}_{H_0}(S \le r_0; \eta,  \vec{q}_1^\eta, \ldots, \vec{q}_{1000}^\eta) \\ 
&=  
\frac{1}{1000}\cdot \sum_{i=1}^{1000}
\bigg( 
\mathbb{I}\{r_i^{(\eta)} \le r_{0} \} \cdot \frac{1}{\eta^K (\prod_{k=1}^K q_{ik}^\eta)^{\eta - 1}}
\bigg)
\end{align*}

We choose $\eta(K, c_t)$ as the root of $\phi(\eta) = c_t$,
which can be numerically obtained using ``uniroot()" function in R.
This choice of $\eta$ guarantees half of the simulated samples will effectively contribute to the importance sampling calculation for each targeted $c_t$.
However, for $c_t \ge 0.01$, we set $\eta = 1$ since the gain of importance sampling diminishes.

\item
\label{step:impS}
(Derive corresponding AW-Fisher statistics for targeted p-value $c_t$)
Next, we derive the corresponding AW-Fisher statistic $S_{K, t}$ for a targeted p-value $c_t$ given $K$.
Given $K$ and $c_t$,
we use $\eta(K, c_t)$ (abbreviated as $\eta$ hereafter) from the previous step to  draw 
$\vec{o}_i = (o_{i1}, \ldots, o_{iK})$,
where $1 \le i \le 10^{7}$ and $\vec{o}_i \overset{i.i.d.}{\sim} \text{Beta}(\eta,1)$.
Denote by $t_i = s(\vec{o}_i)$ the corresponding AW-Fisher statistic of $\vec{o}_i$
and $t_{(1)} \le t_{(2)} \le \ldots \le t_{(10^7)}$ are ordered from $t_1, \ldots, t_{10^7}$.
Define 
\begin{align*} 
m_i &=  \hat{\mathbbm{P}}_{H_0}(S \le t_{(i)}; \eta,  \vec{o}_1, \ldots, \vec{o}_{10^7}) \\ 
&=  
\frac{1}{10^7}\cdot \sum_{j=1}^{10^7}
\bigg( 
\mathbb{I} \{ t_j \le t_{(i)} \} \cdot \frac{1}{\eta^K (\prod_{k=1}^K o_{jk})^{\eta - 1}}
\bigg)
\end{align*}

Note that $m_i$ is monotonically decreasing with $m_1 = 1$ and $m_{10^7} \approx  0 $.
There exists $i^*$ such that $m_{i^*} \le c_t < m_{i^*+1}$.
The corresponding AW-Fisher statistic $S_{K,t}$ given $K$ and $c_t$ is chosen as $S_{K,t} = t(i^*)$.

\item (Interpolation to calculate p-value of a given $S_{obs}$)
From Step~\ref{step:impS},
the library of $c_t$ and $S_{K, t}$ ($t = 1, \ldots, 198$ and $K = 2,\ldots, 100$) is established for interpolation.
For any given AW-Fisher statistic $S_{obs}$ and $K$,
we apply function ``splinefun" in R with ``monoH.FC" option  using $(\log(S_{K,t}), \log(c_t))$, where $t = 1,2,\ldots, 198$,
to fit a smooth curve and identify the corresponding p-value of $S_{obs}$.
Note that we apply spline on log-scale p-value to avoid numerical overflow.
\end{enumerate}

\begin{remark}
In Step~\ref{step:eta}, given $K$,
we simulate $q_{ik} \overset{i.i.d.}{\sim} \text{Unif}(0,1)$ and take the power of $\eta$,
instead of simulating from $\text{Beta}(\eta,1)$.
This design guarantees $\phi(\eta)$ is a monotone function with respect to $\eta$ by eliminating the uncertainty from
sampling $q_{ik}$ for each $\eta$.
\end{remark}

For any future input p-values,
we only need to calculate the AW-Fisher statistics and interpolate the statistics to obtain AW-Fisher p-value by the spline curve fitting.
The design of our base library  $\{(\log(S_{K,t}), \log(c_t))$;  $(t = 1, \ldots, 198,$ and $K = 2,\ldots, 100)\}$ facilitates accurate estimation for AW-Fisher p-value up to precision of $10^{-100}$ and $K$ up to 100.
Although the computation is demanding to generate the base library,
it only runs once before we generate our AW-Fisher R package and will not affect computing for users.
In fact,
it took 6373.5 CPU hours using AMD Opteron(tm) Processor (1.4GHz) to accomplish the whole base library with $10^7$ samples for all $K's$ and $t's$.

\subsection{Simulation and numerical evaluation}
\label{sec:simu}

In section~\ref{sec:pValueComputation} we introduced fast computing for AW-Fisher p-value via importance sampling and interpolation by spline smoothing.
In this section, this interpolation approach will be compared  to the original permutation-based approach in \cite{li2011adaptively} and \cite{wang2012r}.
The comparisons include evaluation of accuracy and computing speed.
In terms of computing  speed, 
our approach applies a new linear sorting algorithm for searching weights and
an interpolation for p-value calculation.
The improvement of linear sorting algorithm is quite obvious:
the searching space reduces from an exponential order $\mathcal{O}(2^K)$ to almost linear order $\mathcal{O}(K \log(K))$.
Below we  utilize the closed-form solution for $K=2$ in Appendix as the underlying truth to compare the new approach with the existing permutation approach.
The linear sorting does not improve computing speed when $K=2$ and the improvement will mainly come from the interpolation.
Our simulation setting is as follows:

\begin{enumerate}
\item Simulate $K = 2$ studies, $G = 10,000$ genes and 2N subjects ($N=20,50$) with  $N$ cases and $N$ controls.
\item 
\label{step:correlatedGenes}
Firstly, we simulated correlated gene structure and assumed no effect size for any gene or any study.
The procedure generally follows \cite{song2014hypothesis}.
\begin{enumerate}
\item
For the first 4,000 genes, simulate 200 gene modules with 20 genes in each module and the remaining 6,000 genes are uncorrelated.
Denote by $C_g \in 
\{
0,1,\ldots,200
\}$
the cluster membership indicator for gene $g$
(e.g. $C_g=1$ indicates gene $g$ is in module 1 while $C_g=0$ indicates gene $g$ is not in any gene module).
\item For module $c$ and study $k$,
simulate $A_{ck}' \sim W^{-1}(\Phi,60)$, 
where $1\le c \le 200$,
$\Phi = 0.5I_{20\times 20} + 0.5J_{20\times 20}$,
$W^{-1}$ denotes the inverse Wishart distribution,
$I$ is the identity matrix and $J$ is the matrix with all elements equal to 1.
$A_{ck}$ is calculated by standardizing $A_{ck}'$ such that the diagonal elements are all 1's.
The covariance matrix for gene module $c$ in study $k$ is calculated as $\Sigma_{ck} = A_{ck}$.
\item
Denote by $g_{c1}, \ldots, g_{c20}$ the indices of the 20 genes in module $c$
(i.e. $C_{g_{cj}} = c$, where $1\le c \le 200$ and $1\le j \le 20$).
Simulate expression levels of genes in module $c$ for sample $n$ in study $k$ as
$(X_{g_{c1}kn}', \ldots, X_{g_{c20}kn}') \sim \mbox{MVN}(0, \Sigma_{cs})$,
where $1\le n \le 2N$ and $1\le k \le K$.
For any uncorrelated gene $g$ with $C_g=0$,
simulate the expression level for sample $n$ in study $k$ as $X_{gkn}' \sim \mbox{N}(0, \sigma^2)$,
where $1\le n \le 2N$ and  $1\le k \le K$.
\end{enumerate}

\item Simulate effect sizes and their DE directions for differentially expressed (DE) genes.

\begin{enumerate}
\item
Assume that the first $G_1$ genes are DE in at least one of the combined studies,
where $G_1 = 30\% \times G$.
For each $1\le g \le G_1$,
simulate $v_g$ from discrete uniform distribution $v_g \sim \mbox{UNIF}(1, \ldots, K)$ 
and then randomly simulate subset $\mathbf{v}_g \subseteq \{1, \ldots, K\}$ such that $|\mathbf{v}_g| = v_g$.
Here $\mathbf{v}_g$ is the set of studies in which gene $g$ is DE.
\item
For any DE gene $g (1\le g\le G_1)$,
simulate gene-level effect size $\theta_g \sim \mbox{N}_{0.5+}(1,1)$,
where $N_{a+}$ denotes the truncated Gaussian distribution within interval $(a,\infty)$.
Also simulate study-specific random effect size $\theta_{gk} \sim \mbox{N} (\theta_g, 0.2^2)$.
\item
Simulate $d_g \sim \mbox{BIN}(1, 0.5)$,
where $1 \le g \le G_1$.
Here $d_g$ is the DE direction for gene $g$ for majority of studies.
\end{enumerate}

\item 
Add the directed effect sizes to the gene expression levels simulated in Step~\ref{step:correlatedGenes}.
For control subjects ($1\le n \le N$), set the expression levels as $X_{gkn} = X_{gkn}'$.
For case subjects $(N+1 \le n \le 2N)$, if $1\le g \le G_1$  and $k \in \mathbf{v}_g$,
we set the expression levels as $X_{gkn} = X_{gkn}' + (-1)^{d_g}\theta_{gk}$.
\end{enumerate}

Using the closed-form solution as the underlying truth,
we evaluated the performance of AW-Fisher p-value from the interpolation approach and the permutation-based approach.
To formally evaluate the accuracy, we utilized root mean square error (rMSE):
$$\mbox{rMSE} = \sqrt{\sum_{g=1}^{G_1} (\alpha_g - \beta_g )^2 / G_1},$$
where $\alpha_g$ is the $-\log_{10}$ (AW-Fisher p-value) for gene $g$ from the permutation approach or the interpolation approach, 
 $\beta_g$ is the $-\log_{10}$ (AW-Fisher p-value) for gene $g$ from closed-form solution
and the rMSE indicates the accuracy of p-value estimates with smaller rMSE for better estimation.
The result for $N=20$ is shown in Table~\ref{tab:deviance} and the result for $N=50$ is in Supplementary Table 1.
Clearly our proposed interpolation approach is superior to permutation-based approach in terms of both accuracy and computing time.
Note that the interpolation approach  is even faster than closed-form solution 
because the interpolation is only based on spline  curve fitting using data in the library and does not 
implement  Monte Carlo importance sampling method 
while the closed form method requires evaluation of power and logarithmic functions.

\begin{table}
\centering
\caption{AW-Fisher p-value accuracy and computing time comparing interpolation approach and permutation-based approach
with $N=20$.
Closed form solution is displayed as benchmark.
$B$ is number of permutations, where closed form solution and interpolation approach don't require any permutation.
Interquartile shows the AW-Fisher p-value range in $-\log_{10}$ scale using closed form solution.
$q_1$ and $q_3$ represent $1^{st}$ and $3^{rd}$ quartile.
rMSE represents root mean squared  error.
}
\label{tab:deviance}
\begin{tabular}{c | c c c c c}
\hline
\hline
Method & B & interquartile($q_1 \sim q_3$) & rMSE & time\\
\hline
closed form & NA & 1.51$ \sim $11.4 & NA & 0.042 secs \\
\cline{1-1}
interpolation & NA & 1.51$ \sim $11.4 & 0.00145 & 0.0115 secs \\
\cline{1-1}
\multirow{3}{*}{permutation}
& 10000 & 1.51$ \sim $11.4 & 10.3 & 2.75 hours \\
& 1000 & 1.51$ \sim $11.4 & 10.8 & 9.52 mins \\
& 100 & 1.51$ \sim $11.4 & 11.4 & 58.9 secs \\
\hline
\end{tabular}
\end{table}%

\section{Variability index of adaptive weights}
\label{sec:variabilityIndex}

\subsection{Method for Variability index}
\label{sec:variabilityIndexProcedure}

As discussed in Issue~\ref{issue:stable} in the Section~\ref{sec:awFisher},
the AW weight estimate $\hat{\textbf{w}}_{g} = (\hat{w}_{g1}, \ldots, \hat{w}_{gK})$ is discontinuous 
as a function of the input p-values and thus may not be stable.
Denote by ${U}_{gk} = 4\cdot\mbox{Var}(\hat{w}_{gk})$
the variability index of AW weight estimate for gene $g$ in study $k$. 
The variability index gauges the stability of $\hat{{w}}_{gk}$, where
a smaller variability index indicates a stable AW weight estimate.
However, ${U}_{gk}$  is not easy to evaluate since $\hat{w}_{gk}$ is binary.
Here, we propose a bootstrap procedure to calculate an estimate of ${U}_{gk}$.
The procedure is as follows:
\begin{enumerate}
\item Obtain a bootstrap sample and repeat the following procedure  $B$ ($b = 1, \ldots, B$) times.
\begin{itemize}
\item Denote by $D_k \in \mathbb{R}^{G\times N_k}$ the data matrix of study $k$,
where $G$ is total number of genes and $N_k$ is total number of samples for study $k$.
$c_{ki}$ is the case-control label, 
where $i \in \{1, \ldots, N_k\}$ is the sample index and
$c_{ki} = 0$ or $1$, representing sample $i$ belongs to control or case group.
\item 
Create an empty data matrix $D_k^{(b)} \in \mathbb{R}^{G\times N_k}$.
Then sample the $i^{th}$ column of $D_k^{(b)}$ using $j^{th}$ column of $D_k$,
where $j \in \{j': c_{kj'} = c_{ki}\}$.
This bootstrap procedure is stepped through for $i = 1, \ldots, N_k$ with replacement (allowing $D_k^{(b)}$
has identical columns).
\item 
Use bootstrapped data matrix $D_k^{(b)}$ to generate AW weight estimate $\hat{w}_{gk}^{(b)}$ and effect size estimate $\hat{\theta}_{gk}^{(b)}$.
\end{itemize}
\item Calculate the variability index estiamte ${\hat{U}}_{gk}$ of $\hat{w}_{gk}$ for gene $g$ in study $k$,
where ${\hat{U}}_{gk} = \frac{4}{B} \sum_{b=1}^B ( \hat{w}_{gk}^{(b)} - \frac{1}{B} \sum_{b'=1}^B  \hat{w}_{gk}^{(b')} )^2$.
\end{enumerate}

Here ${\hat{U}}_{gk}$ ranges from 0 to $1$ with
${\hat{U}}_{gk} = 0$ represents $\hat{w}_{gk}^{(b)} = \hat{w}_{gk}$ for all $b$, 
which indicates stable estimate of AW weight.
${\hat{U}}_{gk} = 1$ represents  $\hat{w}_{gk}^{(b)} = 0$ for half of $b's$ and $\hat{w}_{gk}^{(b)} = 1$ for the other half of $b's$.
A large variability index indicates an  unstable estimate of AW weight.

\subsection{Simulation result}
\label{sec:simuprocedure}
\begin{figure}[htbp]
	\centering
		\includegraphics[height=0.5\columnwidth]{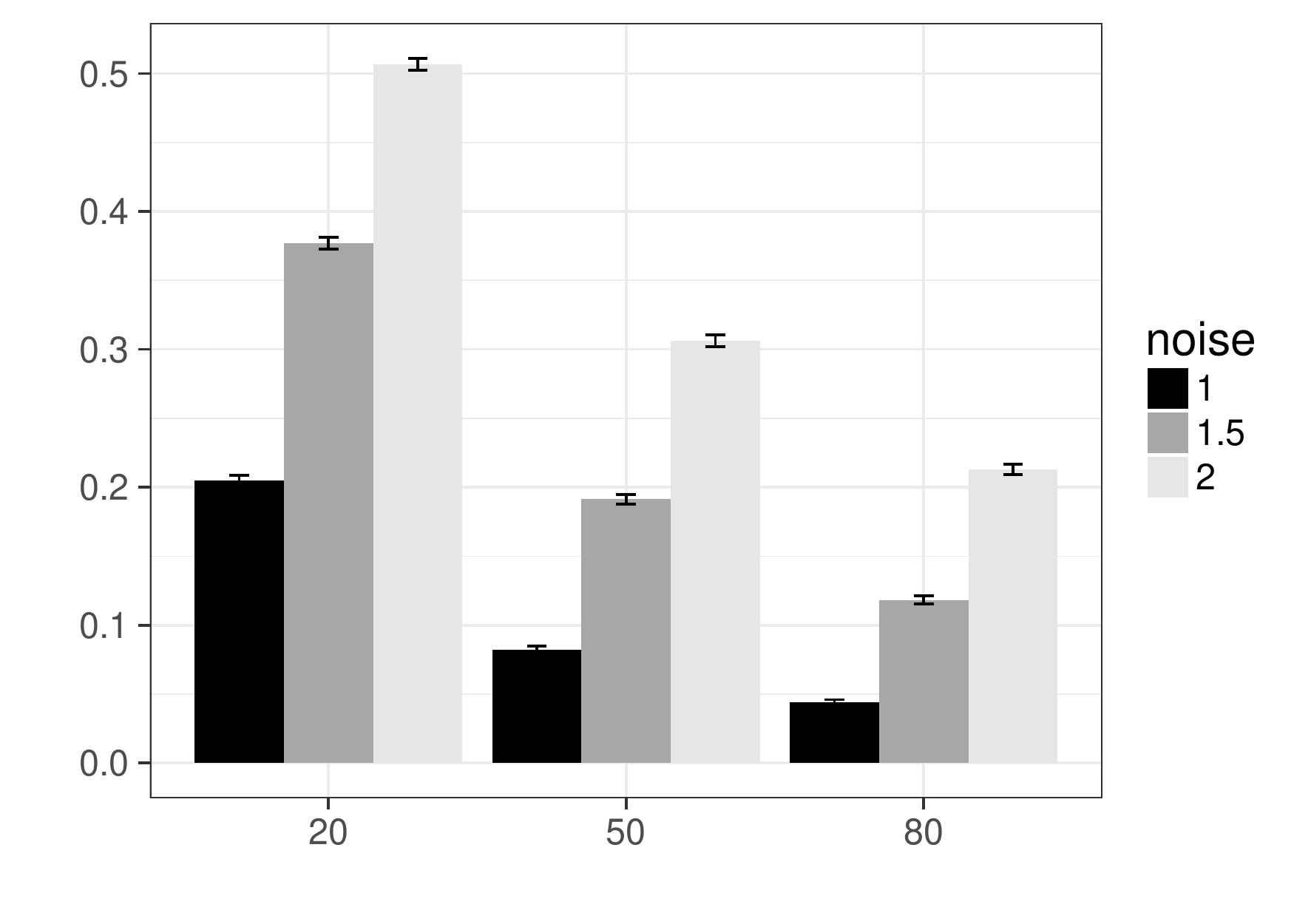}
	\caption{
	Comparison table of variability index for different scenario (combination of sample size and biological variance).
	Only differential expressed genes counting from each individual studies are considered.
	Height of each bar indicates the mean level of variability index and error bar indicates the standard error.
	}
	\label{fig:varibility}
\end{figure}

We followed the simulation setting in Section~\ref{sec:simu} 
to evaluate different combinations of biological variance ($\sigma = 1, 1.5, 2$) and 
sample sizes ($N = 20, 50, 80$) for the performance of the variability index in Figure~\ref{fig:varibility}.
The result shows that when the dataset has smaller sample size or larger biological variation,
the variability index becomes larger.
Since the variability index gauges the stability of AW weight estimate,
it can be seen that noisy datasets tend to generate large variability index.
Back to the two Affymetrix probes shown in Issue~\ref{issue:stable} of Section~\ref{sec:awFisher},
the variability index of $\hat{w} = (1,1,1)$ in $1419484\_a\_at$ is (0, 0.932, 0) and variability index of
 $\hat{w} = (1,0,1)$ in $1425567\_a\_at$ is (0,0.936,0),
 showing unstable weight estimate of the second study for both gene probes.

\section{Resampling-based ensemble clustering for biomarker categorization}
\label{sec:clusters}

\subsection{Method  for biomarker categorization}

In order to categorize detected genes into biomarker groups with similar differential meta-pattern (Issue~\ref{issue:cluster}), 
we extended the bootstrapping procedure in Section~\ref{sec:variabilityIndexProcedure}
to obtain a co-membership matrix for all pairs of genes where each element of the co-membership matrix
represents a similarity of signed AW weight $\hat{\textbf{v}}$ of two genes.
Specifically, denote by $\hat{v}_{gk}^{(b)} = \hat{w}_{gk}^{(b)} \cdot \sgn (\hat{\theta}_{gk}^{(b)})$ from Section~\ref{sec:variabilityIndexProcedure}. 
Define co-membership matrix from each bootstrap sample $b$ as
$W^{(b)}\in \mathbb{R}^{G\times G}$
with elements $W^{(b)}_{gg'} = 1$ if  $\hat{v}_{gk}^{(b)} = \hat{v}_{g'k}^{(b)}$ for all $k$,
and $W^{(b)}_{gg'} = 0$ otherwise. 
The final co-membership matrix is defined as $V = \sum_{b=1}^B W^{(b)} / B$.
We further applied tight clustering algorithm \citep{tseng2005tight}  (``tight.clust" function within R package ``tightClust") to the co-membership matrix $V$ to obtain tight modules.
Tight clustering is able to produce tight and stable gene modules without forcing all genes into clusters.
The resulting gene modules show unique differentially expressed patterns across multiple studies (namely meta-pattern).
We perform the biomarker categorization (clustering) procedure only on declared DE genes at certain false discovery rate cutoff.
Genes of each resulting  module are then sorted by the variability index and visualized by heatmaps.
Below we perform simulation to demonstrate performance of the resampling-based ensemble clustering for biomarker categorization.

\subsection{Simulation result for biomarker categorization}
	To evaluate the performance of biomarker categorization,
	we adopted a simulation procedure similar to Section \ref{sec:simuprocedure} and \cite{huo2017bayesian}.  
	We simulated $S=4$ studies in total and 50 control subjects and 50 case subjects in each study.
	Among the $G = 10,000$ genes, we set 
	$4\%$ as homogeneously concordant DE genes, 
	differentially expressed with the same direction  in all studies (all positive or all negative).
	We denote ``homo$+$'' as the homogeneously concordant DE genes with all positive effect sizes and 
	``homo$-$'' as the homogeneously concordant DE genes with all negative effect sizes.
	We also set another $4\%$ as study-specific DE genes - differential expressed only in one study.
	Among them, 
	$1/4$ are DE genes only in the first study with positive effect sizes (denoted as ``ssp$1+$''),
	$1/4$ are DE genes only in the first study with negative effect sizes (denoted as ``ssp$1-$''),
	$1/4$ are DE genes only in the second study with positive effect sizes (denoted as ``ssp$2+$''),
	and the rest $1/4$ are DE genes only in the second study with negative effect sizes (denoted as ``ssp$2-$'').
	The rest of the genes are non-DE (denoted as ``nonDE'').
	The biological variation parameter  $\sigma$ is set to $1$ in this simulation.

We first applied the proposed AW-Fisher method to this simulated dataset. 
We obtained 794 genes based on  FDR at 5\% under $\mbox{HS}_B$. 
Co-membership of these genes were calculated with $B=1,000$ and used as input for our gene module detection using tight clustering algorithm. 
We identified 6 gene modules in these 794 genes. 
The detected gene modules are tabulated against the true gene modules simulated in Table~\ref{tab:simu2_tight} (Module
0 contains scattered genes not assigned to any of the six modules). 
The false discovery rate is well controlled at $34/794 = 4.3\%$ while the nominal FDR is $5\%$. 
The detected gene modules clearly correspond to the true modules, 
and most of the nonDE genes were left out as  the noises. 
The meta-pattern, variability index and AW weight estimates of these 6 modules are shown in Supplementary Figure~1. 
This simulation study showed that the proposed algorithm can recover the underlying gene meta-pattern.

\begin{table}
\centering
\caption{Contingency table of 794 detected DE genes with simulation underlying truth (on the columns)
and tight clustering result with 6 target modules (on the rows). 
0 represents the scattered gene group.
1 $\sim$ 6 represent 6 detected modules.
Bolded numbers are genes with correct assignment.
}
\label{tab:simu2_tight}
\begin{tabular}{c|ccccccc}
\hline
\hline
Module & homo$-$ & homo$+$ & ssp$1-$ & ssp$1+$ & ssp$2-$ & ssp$2+$ & nonDE\\
\hline
1 & 0 & \textbf{177} & 0 & 0 & 0 & 0 & 0\\
2 & \textbf{184} & 0 & 0 & 0 & 0 & 0 & 0\\
3 & 0 & 0 & 0 & \textbf{74} & 0 & 0 & 1\\
4 & 0 & 0 & \textbf{60} & 0 & 0 & 0 & 1\\
5 & 0 & 0 & 0 & 0 & 0 & \textbf{102} & 2\\
6 & 0 & 0 & 0 & 0 & \textbf{85} & 0 & 3\\
0 & 13 & 24 & 19 & 11 & 6 & 5 & \textbf{27}\\

\hline
\end{tabular}

\end{table}%

\section{Transcriptomic meta-analysis applications}
\label{sec:real}

We applied our proposed methods on two real meta-analysis examples.
The first example utilized gene expression of multi-tissue microarray studies with metabolism related knockout mice.
The second example utilized multi-brain-region RNA-seq studies with HIV transgenic rats.
The sample sizes are shown in Supplementary Table~2.

\subsection{Mouse metabolism example}
\label{sec:metabolism}

Very long-chain acyl-CoA dehydrogenase (VLCAD) deficiency was found to be associated with
energy metabolism disorder in children.
Two genotypes of the mouse model - wild type (VLCAD +/+) and VLCAD-deficient (VLCAD -/-) -
were studied for three types of tissues (brown fat, liver and heart) with 3 to 4 mice in each genotype group.
Total number of probesets from these three transcriptomic microarray studies is 14,495.
Supplementary Table~2a shows details of the study design and the data set is available in supplementary materials.
Two-sided p-values and effect size were calculated using Limma comparing wild type (VLCAD +/+) versus mutant (VLCAD -/-) mice 
in each tissue. 
AW-Fisher meta-analysis p-values were obtained and q-values were calculated  by applying Benjamini-Hochberg procedure.
By controlling FDR at 5\%, we obtained 967 differentially expressed genes.
We calculated the variability index and generated gene co-membership matrix using resampling techniques.
We further applied tight clustering algorithm on the co-membership matrix to identify gene modules with unique meta-pattern.
In this example,
we successfully detected 6 gene modules with different meta-patterns in Figure~\ref{fig:metabolism_modules}.
For example, 
the first and second biomarker modules (gene cluster \Rmnum{1} and \Rmnum{2}) 
are concordant  genes that are up-regulated (or down-regulated) in all tissues.
The other biomarker modules have study-specific differential patterns.
For example, 
DE genes in gene module \Rmnum{3} are up-regulated in heart but not in brown fat or liver.
	To examine the biological functions of these modules, 
	we performed pathway enrichment analysis for genes in each module using Fisher's exact test.
	The pathway database was downloaded from Molecular Signatures Database (MSigDB) v5.0 (\url{http://bioinf.wehi.edu.au/software/MSigDB/}),
	where a mouse-version pathway database were created by combining pathways from KEGG, BIOCARTA, REACTOME and GO databases and mapping all the human genes to their orthologs in mouse
	using Jackson Laboratory Human and Mouse Orthology Report 
	(\url{http://www.informatics.jax.org/orthology.shtml}).
	We summarized the pathway detection result 
	(see supplementary Excel file 1 for detailed pathway information).
	Among the six gene modules with distinct meta-patterns, 
	module \Rmnum{1} is enriched in enzyme activities  (e.g. GO COFACTOR BINDING; $p=3.85\times 10^{-4}$);
	module \Rmnum{2} is enriched in pathways for amino acid catabolism 
	(e.g. REACTOME BRANCHED CHAIN AMINO ACID CATABOLISM; $p=9.31\times 10^{-5}$);
	module \Rmnum{3} is enriched in defense related pathways
	(e.g. DEFENSE RESPONSE;  $p=2.11\times 10^{-6}$);
	module \Rmnum{4} is enriched in pathways of metabolism of amino acids
	(e.g. REACTOME METABOLISM OF AMINO ACIDS; $p=2.36\times 10^{-3}$);
	module \Rmnum{5} is enriched in stimulus related pathways
	(e.g. EXTERNAL STIMULUS;  $p=1.33\times 10^{-3}$);
	For module \Rmnum{6}, we did not detect any significantly  enriched pathways.
	Interestingly, all of these pathways are known to be related to different aspects of metabolism, 
	which indicates that our method is able to detect homogeneous and heterogeneous gene modules that are biologically meaningful.
	The biomarker clustering result enhances meta-analysis interpretation and 
	motivates hypothesis for further biological investigation.
	For example, it is intriguing why the defense  related genes in module \Rmnum{3} are up-regulated only in heart but not in liver and brown fat, 
	and why the stimulus related genes in module \Rmnum{5} are down-regulated in heart and liver but not in brown fat.

\begin{figure}[htbp]
	\centering
	\includegraphics[height=1\columnwidth]{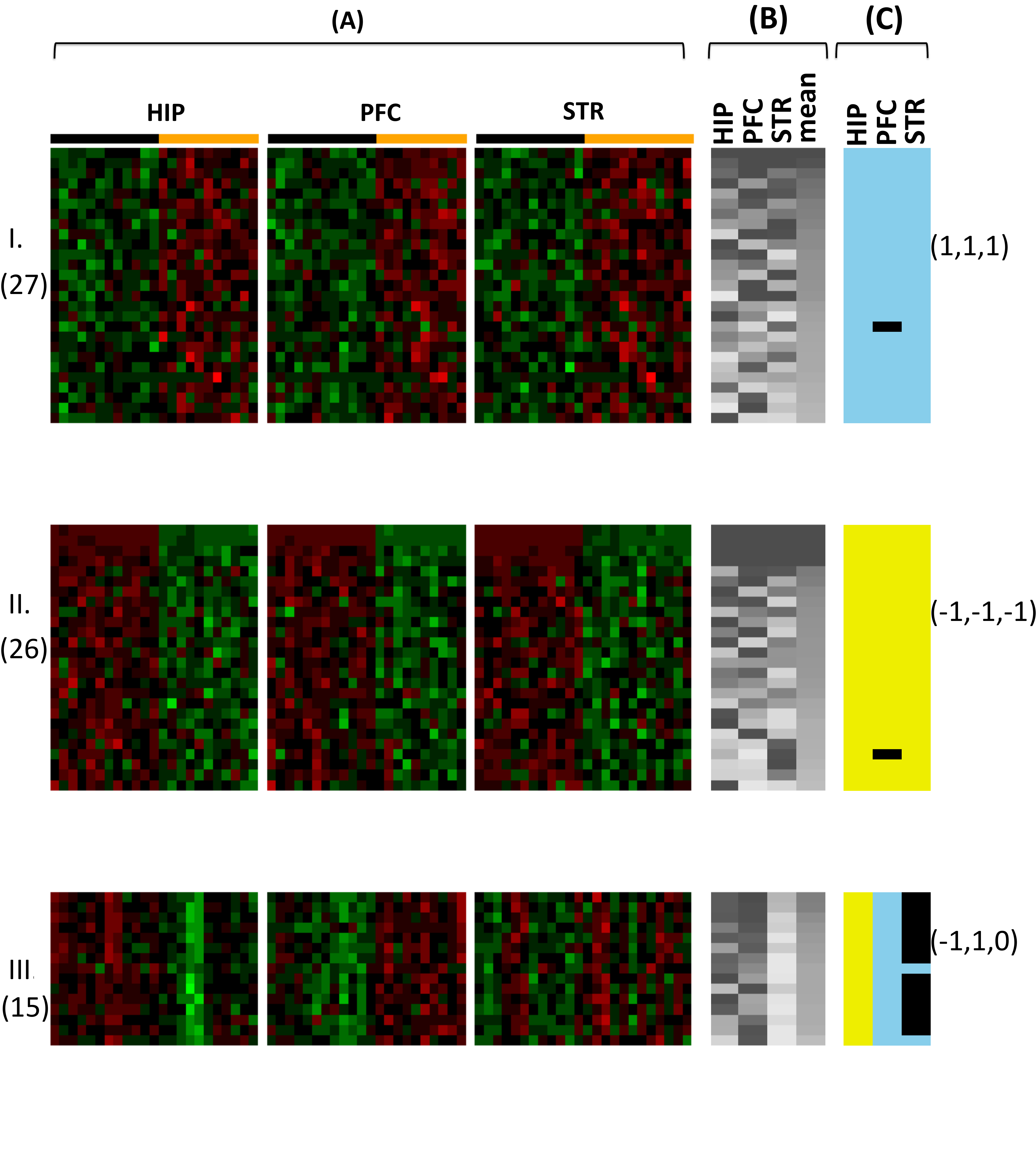}
	\caption{
	Three meta-pattern modules of biomarkers from HIV transgenic rats example.
	Each gene module (Module \Rmnum{1}, \Rmnum{2} and \Rmnum{3}) shows a set of detected biomarkers with similar meta-pattern of differential signals.
	(A) Heatmaps of detected genes (on the rows) and samples (on the columns) for each brain region (HIP, PFC or STR).
		where each brain region represents a study $k$.
	Black color bar on top represents F334 rats (control) and orange color bar on top represents HIV transgenic rats (case).
	Number of genes is shown on the left under each module number. 
	(B) Variability index (genes on the rows and studies on the columns). 
	Variability index is described in Section~\ref{sec:variabilityIndex}, 
	Gene modules, gray heatmap range from 0 (black) to 1 (white), which is the maximum of the variability index.
	Genes of each module are sorted based on the mean variability index.
	(C) AW weight result. 
	Light blue color represents AW weight 1 and up-regulation. 
	Yellow color represents AW weight 1 and down-regulation. 
	Black color represents AW weight 0.
	Genes are shown on the rows and studies are shown on the columns
	Number of genes is shown on right of each module.	}

	\label{fig:HIV_modules}
\end{figure}

\subsection{HIV transgenic rat RNA-seq data}
\cite{li2013transcriptome} conducted studies to determine gene expression differences between F344 and HIV transgenic 
rats using RNA-seq (GSE47474 in Gene expression Omnibus database 
\url{http://www.ncbi.nlm.nih.gov/geo/query/acc.cgi?acc=GSE47474}).
The HIV transgenic rat model is designed to study learning, memory, 
vulnerability to drug addiction and other psychiatric disorders to HIV positive patients.
12 F334 untreated rats and 12 HIV transgenic rats in prefrontal cortex (PFC), hippocampus (HIP), and striatum (STR) regions are sequenced for RNA-seq
(see Supplementary Table~2b. 
Tophat \citep{trapnell2009tophat} was applied for alignment (adopted by \cite{li2013transcriptome}) and the alignment results were  converted to RNA-seq count data with 16,821 genes by bedtools \citep{quinlan2010bedtools}.
Genes with less than 100 total counts within any brain region were filtered out and 11,824 genes remained.
Potential outliers were removed by checking the sample correlation heatmaps (see Supplementary Figure~2). 
R package ``edgeR" \citep{robinson2010edger} was adopted to perform differential expression gene detection and two-sided p-values were obtained.
AW-Fisher meta-analysis p-values were evaluated and q-values were obtained by applying Benjamini-Hochberg procedure.
By controlling FDR at 30\%, we obtained 145 differentially expressed genes.
We loose the FDR criteria to  30\% since it is well known that the transcriptomic signals in brain are generally weak.
We calculated the variability index and performed biomarker categorization by using resampling techniques and tight clustering algorithm.
The result is shown in  Figure~\ref{fig:HIV_modules}.
To examine the biological functions of these modules, 
we also performed pathway enrichment analysis using the same procedure as in Section~\ref{sec:metabolism}
(see supplementary Excel file 2 for detailed information).  
As the results show, module \Rmnum{1} is up-regulated in all the three brain regions, 
and is enriched in pathways related to response to virus.
(e.g. GO RESPONSE TO VIRUS; $p = 1.59 \times 10^{-3}$); 
module \Rmnum{2} is down-regulated in all the three brain regions, 
and is enriched in pathways related to rhythmic process
(e.g. GO RHYTHMIC PROCESS; $p = 6.23 \times 10^{-4}$); 
module \Rmnum{3} is especially interesting since it is down-regulated in HIP, 
but up-regulated in PFC and STR. 
However we did not detect any significant pathways using MSigDB,
possibly  due to small module size (only 15 genes).
Instead we used a broader mouse pathway database from Gene Ontology Consortium \citep{Bares2015gskb},
which contained broader pathway categories;
we found that GO FOREBRAIN DEVELOPMENT and GO TELENCEPHALON DEVELOPMENT pathways are highly associated with module \Rmnum{3} ($p = 2.82 \times 10^{-3}$ and $p = 2.84 \times 10^{-4}$).
Since the brain regions are affected by virus, 
we anticipate that genes responding to virus to be up-regulated, as observed in module \Rmnum{1}.
The down-regulation of rhythmic process genes in module \Rmnum{2}  indicates that HIV virus may have caused loss of rhythmic pattern in multiple brain regions.
Moreover, because different brain regions have different functions, 
it is not surprising that some brain development related genes (module \Rmnum{3}) respond differently to HIV in different brain regions.

\label{sec:hiv}

\section{Conclusion and discussion}
\label{sec:conclusion}

Emerging omics datasets in public domain has made genome-wide meta-analysis appealing. 
Adaptively weighted Fisher's method has become useful and popular in the stance that 
it will characterizes study-specific contributions  to the meta-analysis result. 
In this  paper, we proposed fast computing and biomarker clustering methods to improve application of  AW-Fisher. 
The contributions of this paper are threefold:  
(1) Previous version of AW algorithm relied on permutation analysis to assess p-values, 
which set a limitation for accuracy and speed. 
We proposed a fast computing and weight searching algorithm for AW algorithm based on 
importance sampling, interpolation and a linear searching complexity of AW weight, 
which makes the AW-Fisher algorithm more applicable for large-scale genomic applications. 
(2) We developed an AW-Fisher weight variability index.
This is essential to determine stability of AW-Fisher weight estimates.
(3) We proposed a biomarker categorization algorithm via a resampling procedure, 
which can efficiently obtain gene modules of different  meta-analysis differential expression pattern (namely meta-patterns). 
These meta-patterns can help establish biological hypothesis to quantify homogeneous and heterogeneous DE signals across studies and guide next-step biological investigation. 
Finally, the the  superior performances of the proposed methods are demonstrated in simulation and two real applications (mouse brain HIV RNA-seq data and mouse metabolism data). 

We note that the adaptive weight concept can be extended from Fisher's method to other p-value combination meta-analysis methods,
such as Stouffer's method.
The linear weight searching, importance sampling and spline smoothing can equally be applied in order to efficiently obtain accurate p-values (e.g. AW-Stouffer's method).
An R package (calling C++) is available \url{https://github.com/Caleb-Huo/AWFisher}
and all datasets and programming code used to perform all analyses  in this paper are available on author's website.

\bibliographystyle{imsart-nameyear}
\bibliography{AWcomputing}

\begin{thebibliography}{30}

\bibitem[\protect\citeauthoryear{Bares and Ge}{2015}]{Bares2015gskb}
\begin{bmanual}[author]
\bauthor{\bsnm{Bares},~\bfnm{Valerie}\binits{V.}} \AND
  \bauthor{\bsnm{Ge},~\bfnm{Xijin}\binits{X.}}
(\byear{2015}).
\btitle{gskb: Gene Set data for pathway analysis in mouse}
\bnote{R package version 1.3.0}.
\end{bmanual}
\endbibitem

\bibitem[\protect\citeauthoryear{Begum et~al.}{2012}]{begum2012comprehensive}
\begin{barticle}[author]
\bauthor{\bsnm{Begum},~\bfnm{Ferdouse}\binits{F.}},
  \bauthor{\bsnm{Ghosh},~\bfnm{Debashis}\binits{D.}},
  \bauthor{\bsnm{Tseng},~\bfnm{George~C}\binits{G.~C.}} \AND
  \bauthor{\bsnm{Feingold},~\bfnm{Eleanor}\binits{E.}}
(\byear{2012}).
\btitle{Comprehensive literature review and statistical considerations for GWAS
  meta-analysis}.
\bjournal{Nucleic acids research}
\bpages{gkr1255}.
\end{barticle}
\endbibitem

\bibitem[\protect\citeauthoryear{Benjamini and
  Heller}{2008}]{benjamini2008screening}
\begin{barticle}[author]
\bauthor{\bsnm{Benjamini},~\bfnm{Yoav}\binits{Y.}} \AND
  \bauthor{\bsnm{Heller},~\bfnm{Ruth}\binits{R.}}
(\byear{2008}).
\btitle{Screening for partial conjunction hypotheses}.
\bjournal{Biometrics}
\bvolume{64}
\bpages{1215--1222}.
\end{barticle}
\endbibitem

\bibitem[\protect\citeauthoryear{Benjamini and
  Hochberg}{1995}]{benjamini1995controlling}
\begin{barticle}[author]
\bauthor{\bsnm{Benjamini},~\bfnm{Y.}\binits{Y.}} \AND
  \bauthor{\bsnm{Hochberg},~\bfnm{Y.}\binits{Y.}}
(\byear{1995}).
\btitle{Controlling the false discovery rate: a practical and powerful approach
  to multiple testing}.
\bjournal{Journal of the Royal Statistical Society. Series B (Methodological)}
\bpages{289--300}.
\end{barticle}
\endbibitem

\bibitem[\protect\citeauthoryear{Birnbaum}{1954}]{birnbaum1954combining}
\begin{barticle}[author]
\bauthor{\bsnm{Birnbaum},~\bfnm{A.}\binits{A.}}
(\byear{1954}).
\btitle{Combining independent tests of significance}.
\bjournal{Journal of the American Statistical Association}
\bpages{559--574}.
\end{barticle}
\endbibitem

\bibitem[\protect\citeauthoryear{Chang et~al.}{2013}]{chang2013meta}
\begin{barticle}[author]
\bauthor{\bsnm{Chang},~\bfnm{Lun-Ching}\binits{L.-C.}},
  \bauthor{\bsnm{Lin},~\bfnm{Hui-Min}\binits{H.-M.}},
  \bauthor{\bsnm{Sibille},~\bfnm{Etienne}\binits{E.}} \AND
  \bauthor{\bsnm{Tseng},~\bfnm{George~C}\binits{G.~C.}}
(\byear{2013}).
\btitle{Meta-analysis methods for combining multiple expression profiles:
  comparisons, statistical characterization and an application guideline}.
\bjournal{BMC bioinformatics}
\bvolume{14}
\bpages{368}.
\end{barticle}
\endbibitem

\bibitem[\protect\citeauthoryear{Domany}{2014}]{domany2014using}
\begin{barticle}[author]
\bauthor{\bsnm{Domany},~\bfnm{Eytan}\binits{E.}}
(\byear{2014}).
\btitle{Using high-throughput transcriptomic data for prognosis: a critical
  overview and perspectives}.
\bjournal{Cancer research}
\bvolume{74}
\bpages{4612--4621}.
\end{barticle}
\endbibitem

\bibitem[\protect\citeauthoryear{Fisher}{1934}]{fisher1934statistical}
\begin{barticle}[author]
\bauthor{\bsnm{Fisher},~\bfnm{Ronald~Aylmer}\binits{R.~A.}}
(\byear{1934}).
\btitle{Statistical methods for research workers}.
\end{barticle}
\endbibitem

\bibitem[\protect\citeauthoryear{Huo, Song and Tseng}{2017}]{huo2017bayesian}
\begin{barticle}[author]
\bauthor{\bsnm{Huo},~\bfnm{Zhiguang}\binits{Z.}},
  \bauthor{\bsnm{Song},~\bfnm{Chi}\binits{C.}} \AND
  \bauthor{\bsnm{Tseng},~\bfnm{George}\binits{G.}}
(\byear{2017}).
\btitle{Bayesian latent hierarchical model for transcriptomic meta-analysis to
  detect biomarkers with clustered meta-patterns of differential expression
  signals}.
\bjournal{arXiv preprint arXiv:1707.03301}.
\end{barticle}
\endbibitem

\bibitem[\protect\citeauthoryear{Li and Tseng}{2011}]{li2011adaptively}
\begin{barticle}[author]
\bauthor{\bsnm{Li},~\bfnm{J.}\binits{J.}} \AND
  \bauthor{\bsnm{Tseng},~\bfnm{G.~C.}\binits{G.~C.}}
(\byear{2011}).
\btitle{An adaptively weighted statistic for detecting differential gene
  expression when combining multiple transcriptomic studies}.
\bjournal{The Annals of Applied Statistics}
\bvolume{5}
\bpages{994--1019}.
\end{barticle}
\endbibitem

\bibitem[\protect\citeauthoryear{Li et~al.}{2013}]{li2013transcriptome}
\begin{barticle}[author]
\bauthor{\bsnm{Li},~\bfnm{Ming~D}\binits{M.~D.}},
  \bauthor{\bsnm{Cao},~\bfnm{Junran}\binits{J.}},
  \bauthor{\bsnm{Wang},~\bfnm{Shaolin}\binits{S.}},
  \bauthor{\bsnm{Wang},~\bfnm{Ju}\binits{J.}},
  \bauthor{\bsnm{Sarkar},~\bfnm{Sraboni}\binits{S.}},
  \bauthor{\bsnm{Vigorito},~\bfnm{Michael}\binits{M.}},
  \bauthor{\bsnm{Ma},~\bfnm{Jennie~Z}\binits{J.~Z.}} \AND
  \bauthor{\bsnm{Chang},~\bfnm{Sulie~L}\binits{S.~L.}}
(\byear{2013}).
\btitle{Transcriptome sequencing of gene expression in the brain of the HIV-1
  transgenic rat}.
\bjournal{PloS one}
\bvolume{8}
\bpages{e59582}.
\end{barticle}
\endbibitem

\bibitem[\protect\citeauthoryear{Littell and
  Folks}{1971}]{littell1971asymptotic}
\begin{barticle}[author]
\bauthor{\bsnm{Littell},~\bfnm{R.~C.}\binits{R.~C.}} \AND
  \bauthor{\bsnm{Folks},~\bfnm{J.~L.}\binits{J.~L.}}
(\byear{1971}).
\btitle{Asymptotic optimality of Fisher's method of combining independent
  tests}.
\bjournal{Journal of the American Statistical Association}
\bpages{802--806}.
\end{barticle}
\endbibitem

\bibitem[\protect\citeauthoryear{Pan}{2002}]{pan2002comparative}
\begin{barticle}[author]
\bauthor{\bsnm{Pan},~\bfnm{Wei}\binits{W.}}
(\byear{2002}).
\btitle{A comparative review of statistical methods for discovering
  differentially expressed genes in replicated microarray experiments}.
\bjournal{Bioinformatics}
\bvolume{18}
\bpages{546--554}.
\end{barticle}
\endbibitem

\bibitem[\protect\citeauthoryear{Park et~al.}{2017+}]{Park2017AWtheory}
\begin{barticle}[author]
\bauthor{\bsnm{Park},~\bfnm{YongSeok}\binits{Y.}},
  \bauthor{\bsnm{Huo},~\bfnm{Zhiguang}\binits{Z.}},
  \bauthor{\bsnm{Tang},~\bfnm{Shaowu}\binits{S.}} \AND
  \bauthor{\bsnm{Tseng},~\bfnm{George}\binits{G.}}
(\byear{2017}+).
\btitle{Asymptotic properties of adaptive weighted Fisher's method}.
\end{barticle}
\endbibitem

\bibitem[\protect\citeauthoryear{Quinlan and Hall}{2010}]{quinlan2010bedtools}
\begin{barticle}[author]
\bauthor{\bsnm{Quinlan},~\bfnm{Aaron~R}\binits{A.~R.}} \AND
  \bauthor{\bsnm{Hall},~\bfnm{Ira~M}\binits{I.~M.}}
(\byear{2010}).
\btitle{BEDTools: a flexible suite of utilities for comparing genomic
  features}.
\bjournal{Bioinformatics}
\bvolume{26}
\bpages{841--842}.
\end{barticle}
\endbibitem

\bibitem[\protect\citeauthoryear{Ramasamy et~al.}{2008}]{ramasamy2008key}
\begin{barticle}[author]
\bauthor{\bsnm{Ramasamy},~\bfnm{Adaikalavan}\binits{A.}},
  \bauthor{\bsnm{Mondry},~\bfnm{Adrian}\binits{A.}},
  \bauthor{\bsnm{Holmes},~\bfnm{Chris~C}\binits{C.~C.}} \AND
  \bauthor{\bsnm{Altman},~\bfnm{Douglas~G}\binits{D.~G.}}
(\byear{2008}).
\btitle{Key issues in conducting a meta-analysis of gene expression microarray
  datasets}.
\bjournal{PLoS Med}
\bvolume{5}
\bpages{e184}.
\end{barticle}
\endbibitem

\bibitem[\protect\citeauthoryear{Robinson, McCarthy and
  Smyth}{2010}]{robinson2010edger}
\begin{barticle}[author]
\bauthor{\bsnm{Robinson},~\bfnm{Mark~D}\binits{M.~D.}},
  \bauthor{\bsnm{McCarthy},~\bfnm{Davis~J}\binits{D.~J.}} \AND
  \bauthor{\bsnm{Smyth},~\bfnm{Gordon~K}\binits{G.~K.}}
(\byear{2010}).
\btitle{edgeR: a Bioconductor package for differential expression analysis of
  digital gene expression data}.
\bjournal{Bioinformatics}
\bvolume{26}
\bpages{139--140}.
\end{barticle}
\endbibitem

\bibitem[\protect\citeauthoryear{Roy}{1953}]{roy1953heuristic}
\begin{barticle}[author]
\bauthor{\bsnm{Roy},~\bfnm{SN}\binits{S.}}
(\byear{1953}).
\btitle{On a heuristic method of test construction and its use in multivariate
  analysis}.
\bjournal{The Annals of Mathematical Statistics}
\bpages{220--238}.
\end{barticle}
\endbibitem

\bibitem[\protect\citeauthoryear{Simon}{2005}]{simon2005development}
\begin{barticle}[author]
\bauthor{\bsnm{Simon},~\bfnm{Richard}\binits{R.}}
(\byear{2005}).
\btitle{Development and validation of therapeutically relevant multi-gene
  biomarker classifiers}.
\bjournal{Journal of the National Cancer Institute}
\bvolume{97}
\bpages{866--867}.
\end{barticle}
\endbibitem

\bibitem[\protect\citeauthoryear{Simon et~al.}{2003}]{simon2003pitfalls}
\begin{barticle}[author]
\bauthor{\bsnm{Simon},~\bfnm{Richard}\binits{R.}},
  \bauthor{\bsnm{Radmacher},~\bfnm{Michael~D}\binits{M.~D.}},
  \bauthor{\bsnm{Dobbin},~\bfnm{Kevin}\binits{K.}} \AND
  \bauthor{\bsnm{McShane},~\bfnm{Lisa~M}\binits{L.~M.}}
(\byear{2003}).
\btitle{Pitfalls in the use of DNA microarray data for diagnostic and
  prognostic classification}.
\bjournal{Journal of the National Cancer Institute}
\bvolume{95}
\bpages{14--18}.
\end{barticle}
\endbibitem

\bibitem[\protect\citeauthoryear{Smyth}{2005}]{smyth2005limma}
\begin{bincollection}[author]
\bauthor{\bsnm{Smyth},~\bfnm{Gordon~K}\binits{G.~K.}}
(\byear{2005}).
\btitle{Limma: linear models for microarray data}.
In \bbooktitle{Bioinformatics and computational biology solutions using R and
  Bioconductor}
\bpages{397--420}.
\bpublisher{Springer}.
\end{bincollection}
\endbibitem

\bibitem[\protect\citeauthoryear{Soneson and
  Delorenzi}{2013}]{soneson2013comparison}
\begin{barticle}[author]
\bauthor{\bsnm{Soneson},~\bfnm{Charlotte}\binits{C.}} \AND
  \bauthor{\bsnm{Delorenzi},~\bfnm{Mauro}\binits{M.}}
(\byear{2013}).
\btitle{A comparison of methods for differential expression analysis of RNA-seq
  data}.
\bjournal{BMC bioinformatics}
\bvolume{14}
\bpages{1}.
\end{barticle}
\endbibitem

\bibitem[\protect\citeauthoryear{Song and Tseng}{2014}]{song2014hypothesis}
\begin{barticle}[author]
\bauthor{\bsnm{Song},~\bfnm{Chi}\binits{C.}} \AND
  \bauthor{\bsnm{Tseng},~\bfnm{George~C}\binits{G.~C.}}
(\byear{2014}).
\btitle{Hypothesis setting and order statistic for robust genomic
  meta-analysis}.
\bjournal{The annals of applied statistics}
\bvolume{8}
\bpages{777}.
\end{barticle}
\endbibitem

\bibitem[\protect\citeauthoryear{Stouffer et~al.}{1949}]{stouffer1949american}
\begin{bbook}[author]
\bauthor{\bsnm{Stouffer},~\bfnm{S.~A.}\binits{S.~A.}},
  \bauthor{\bsnm{Suchman},~\bfnm{E.~A.}\binits{E.~A.}},
  \bauthor{\bsnm{Devinney},~\bfnm{L.~C.}\binits{L.~C.}},
  \bauthor{\bsnm{Star},~\bfnm{S.~A.}\binits{S.~A.}} \AND
  \bauthor{\bsnm{Williams~Jr},~\bfnm{R.~M.}\binits{R.~M.}}
(\byear{1949}).
\btitle{The American soldier: adjustment during army life.}
\bpublisher{Princeton Univ. Press}.
\end{bbook}
\endbibitem

\bibitem[\protect\citeauthoryear{Sun and Wright}{2010}]{sun2010geometric}
\begin{barticle}[author]
\bauthor{\bsnm{Sun},~\bfnm{Wei}\binits{W.}} \AND
  \bauthor{\bsnm{Wright},~\bfnm{Fred~A}\binits{F.~A.}}
(\byear{2010}).
\btitle{A geometric interpretation of the permutation p-value and its
  application in eQTL studies}.
\bjournal{The Annals of Applied Statistics}
\bpages{1014--1033}.
\end{barticle}
\endbibitem

\bibitem[\protect\citeauthoryear{Tippett}{1931}]{tippett1931methods}
\begin{bbook}[author]
\bauthor{\bsnm{Tippett},~\bfnm{L.~H.~C.}\binits{L.~H.~C.}}
(\byear{1931}).
\btitle{The Methods of Statistics.}
\bpublisher{London: Williams Norgate Ltd.}
\end{bbook}
\endbibitem

\bibitem[\protect\citeauthoryear{Trapnell, Pachter and
  Salzberg}{2009}]{trapnell2009tophat}
\begin{barticle}[author]
\bauthor{\bsnm{Trapnell},~\bfnm{Cole}\binits{C.}},
  \bauthor{\bsnm{Pachter},~\bfnm{Lior}\binits{L.}} \AND
  \bauthor{\bsnm{Salzberg},~\bfnm{Steven~L}\binits{S.~L.}}
(\byear{2009}).
\btitle{TopHat: discovering splice junctions with RNA-Seq}.
\bjournal{Bioinformatics}
\bvolume{25}
\bpages{1105--1111}.
\end{barticle}
\endbibitem

\bibitem[\protect\citeauthoryear{Tseng, Ghosh and
  Feingold}{2012}]{tseng2012comprehensive}
\begin{barticle}[author]
\bauthor{\bsnm{Tseng},~\bfnm{G.~C.}\binits{G.~C.}},
  \bauthor{\bsnm{Ghosh},~\bfnm{D.}\binits{D.}} \AND
  \bauthor{\bsnm{Feingold},~\bfnm{E.}\binits{E.}}
(\byear{2012}).
\btitle{Comprehensive literature review and statistical considerations for
  microarray meta-analysis}.
\bjournal{Nucleic Acids Research}.
\end{barticle}
\endbibitem

\bibitem[\protect\citeauthoryear{Tseng and Wong}{2005}]{tseng2005tight}
\begin{barticle}[author]
\bauthor{\bsnm{Tseng},~\bfnm{George~C}\binits{G.~C.}} \AND
  \bauthor{\bsnm{Wong},~\bfnm{Wing~H}\binits{W.~H.}}
(\byear{2005}).
\btitle{Tight Clustering: A Resampling-Based Approach for Identifying Stable
  and Tight Patterns in Data}.
\bjournal{Biometrics}
\bvolume{61}
\bpages{10--16}.
\end{barticle}
\endbibitem

\bibitem[\protect\citeauthoryear{Wang et~al.}{2012}]{wang2012r}
\begin{barticle}[author]
\bauthor{\bsnm{Wang},~\bfnm{Xingbin}\binits{X.}},
  \bauthor{\bsnm{Kang},~\bfnm{Dongwan~D}\binits{D.~D.}},
  \bauthor{\bsnm{Shen},~\bfnm{Kui}\binits{K.}},
  \bauthor{\bsnm{Song},~\bfnm{Chi}\binits{C.}},
  \bauthor{\bsnm{Lu},~\bfnm{Shuya}\binits{S.}},
  \bauthor{\bsnm{Chang},~\bfnm{Lun-Ching}\binits{L.-C.}},
  \bauthor{\bsnm{Liao},~\bfnm{Serena~G}\binits{S.~G.}},
  \bauthor{\bsnm{Huo},~\bfnm{Zhiguang}\binits{Z.}},
  \bauthor{\bsnm{Tang},~\bfnm{Shaowu}\binits{S.}},
  \bauthor{\bsnm{Ding},~\bfnm{Ying}\binits{Y.}} \betal{et~al.}
(\byear{2012}).
\btitle{An R package suite for microarray meta-analysis in quality control,
  differentially expressed gene analysis and pathway enrichment detection}.
\bjournal{Bioinformatics}
\bvolume{28}
\bpages{2534--2536}.
\end{barticle}
\endbibitem

\end{thebibliography}

\end{document}